\documentclass[12pt]{article}
\usepackage{epsfig}
\begin{document}
\begin{center}
{\Large {\bf 10 D Euclidean dynamical triangulations.  }}

\vskip-40mm \rightline{\small ITEP-LAT/2003-31 } \vskip 30mm

{
\vspace{1cm}
{A.I.~Veselov, M.A.~Zubkov  }\\
\vspace{.5cm} { \it ITEP, B.Cheremushkinskaya 25, Moscow, 117259, Russia }}
\end{center}
\begin{abstract}
We investigate numerically $10$ - dimensional Euclidean quantum gravity (with
discretized Einstein - Hilbert action) in the framework of the dynamical
triangulation approach. For the considered values of the gravitational coupling
we observed two phases, the behavior of which is found to be similar to that of
the crumpled and elongated phases of $3$, $4$ and $5$ dimensional models.
Surprisingly, (for the observed lattice sizes) the natural state of the $10$ D
system (when the Einstein - Hilbert action is turned off) is found to resemble
branched polymer while in the low dimensional systems the natural state belongs
to the crumpled phase.
\end{abstract}


\newpage

\section{Introduction}
The attempts to unify fundamental interactions have produced a lot of
  various models (see, for example, \cite{Strings}
  and references therein). All of them are of rich
mathematical structure and most of them are to some degree
 based on the Riemannian geometry. That's why we believe that the
 quantization of the
 latter is rather important. Even if it has nothing to do with
 the real gravity, it may play an important role
 in the further construction of the realistic unified theory.

Recently it has been paid much attention to the dynamical triangulation (DT)
 approach to quantization of Riemannian geometry \cite{Triangulation} in
 $2$, $3$ and $4$ dimensions\footnote{$5$ - dimensional Euclidean quantum gravity
 was considered in \cite{5D}.}. For $D = 2$ the DT model
  has a well - defined
 continuum limit consistent with the predictions of the continuum theory
  \cite{2D}. At $D = 3,4$ the Euclidean DT models with the discretized
  Einstein action have two phases: the crumpled phase with infinite fractal dimension
   and the elongated one, which resembles branched polymer model with the
 fractal dimension close to $2$. It appears that the introducing of the causal structure to these
 models (correspondent to the transition from Euclidean to Lorentzian
 quantum gravity) or coupling them to matter changes their behavior and make them more
 realistic \cite{Lorentzian,Gauge_Triang}. Nevertheless,  the pure
Euclidean gravity is still of interest as an area of developing the
investigation methods. Moreover, mechanisms observed within this model can be
related to more realistic models.

The mathematical structures related to the unification of fundamental forces
may include the concept of high dimensional ($D > 4$) space. In particular, all
known superstring models become unambiguous only in the  $10$ - dimensional
space - time. Unfortunately, most of information about their structure comes
from the perturbative methods and the (incomplete) investigation of certain
special excitations (such as $D$ - branes). It seems that the nonperturbative
investigation of those models would become exhaustive only if the numerical
lattice methods are used. The modern lattice theory has an experience of
dealing with quantum gravity models in lower dimensions. In particular,  the
dynamical triangulation method was applied to the systems of $D = 2,3,4$
dimensions. Therefore we guess that it is reasonable to apply this method to
the higher dimensional problems. The investigation of the pure Einstein
Euclidean gravity can become
 the first step. The work related to this step is the content of
the present paper.

Namely, we consider the ten - dimensional Euclidean dynamical triangulations of
spherical topology. Their behavior was expected to be similar to that of $3,4$
and $5$ - dimensional systems. This supposition was confirmed partially.
However, it turns out that there are a few features that are not present in the
lower dimensional models. For example, the $10D$ model  does have two phases
that are similar to the crumpled and elongated phases of $3$, $4$ and $5$ -
dimensional models. But in our case the phase transition (at least for the
observed volumes $V = 8000$ and $V = 32000$) is at negative gravitational
coupling constant. This means that at "physical" positive gravitational
constant $G$ the considered $10$ - dimensional system cannot exist in the
crumpled phase unlike $3$, $4$ and $5$ - dimensional systems. Also this means
that for the observed lattice sizes the natural state of the $10$ D system
(when the Einstein - Hilbert action is turned off) resembles branched polymer
while in the low dimensional systems the natural state belongs to the crumpled
phase.

It is worth mentioning that the considered volumes  are not consistent with the
system of "physical" dimension $D = 10$. It follows from the observation that
 already at linear size $3$ the analogous rectangular $10D$ lattice has the
volume $V = 3^{10} = 59049$. However, the observed effective (fractal)
dimension of the triangulated space, say,  at $V = 32000$ is $\sim 4$ in the
crumpled phase and $\sim 2$ in the elongated one. So, the linear size of the
system is expected to be of the order of $\sim (32000)^{\frac{1}{4}}\sim 10$
and $\sim (32000)^{\frac{1}{2}} \sim 100 $ respectively. This is confirmed by
the direct measurement of the linear extent.

\section{The model.}

In this section we shall remind briefly the definition of the considered model
and the description of the numerical algorithm. For the complete review of the
dynamical triangulations see \cite{Triangulation} and references therein. For
the full description of the numerical algorithm for arbitrary dimension see
\cite{Catterall}.

In the dynamical triangulation approach the Riemannian manifold (of Euclidian
signature) is approximated by the simplicial complex obtained by gluing the $D$
- dimensional simplices. Each simplex has $D+1$ vertices. All links are assumed
to have the same length $a$. The metrics inside each simplex is supposed to be
flat. Therefore the deviation from the flatness is concentrated on the
boundaries of the simplices. The scalar curvature $R$ is zero everywhere except
the bones ($D-2$ - dimensional subsimplices of the triangulation simplices).
The Einstein - Hilbert action can be expressed through the number of bones and
the number of simplices of the given triangulated manyfold:
\begin{eqnarray}
S & = & - \frac{1}{16 \pi G} \int R(x) \sqrt{g} d^D x =  - \frac{{\rm
Vol}_{D-2}}{16 \pi G}\sum_{\rm bones} (2\pi - O({\rm bone}) cos^{-1}
(\frac{1}{D}))
\nonumber\\
 & = & - \frac{{\rm Vol}_{D-2}}{8 G} ( N_{\rm bones} -
 \frac{D(D+1)}{4\pi} N_{\rm simplices} cos^{-1} (\frac{1}{D})),
\end{eqnarray}
where $O({\rm bone})$ is the number of simplices sharing the given bone, ${\rm
Vol}_{j} = \frac{a^j \sqrt{j+1}}{j!\sqrt{2^j}}$ is the volume of a $j$ -
dimensional simplex with the edges of length $a$, $N_{\rm bones}$ is the total
number of bones and $N_{\rm simplices}$ is the total number of simplices.

The metric of the triangulated manyfold is completely defined by the way the
simplices are glued together. Therefore, the functional integral over $D g$ in
this approach is changed by the summation over the different triangulations (we
restrict ourselves with the spherical topology only):
\begin{equation}
\int D g \rightarrow \sum_{T}\frac{1}{C_T}
\end{equation}
Where the sum is over the triangulations $T$ that approximate different
Riemannian manyfolds\footnote{Two formally different triangulations may
approximate the same Riemannian manyfold. Therefore the correspondent
factorization should be implemented.} and $C_T$ is the symmetry factor of the
triangulation itself (the order of its automorphism group).

We consider the model, in which the fluctuations of the global invariant $D$ -
volume are suppressed. The partition function has the form:
\begin{equation}
Z_V = \sum_T \frac{1}{C_T}{\rm exp} ( - S(T)) = \sum_T \frac{1}{C_T}{\rm
exp}(\kappa_{D - 2} N_{D - 2} - \kappa_{D} N_D - \gamma (N_D - V)^2
)\label{ZVD}
\end{equation}
where we denoted $N_D = N_{\rm simplices}$ , $N_{D-2} = N_{\rm bones}$ , and
$\kappa_{D-2} = \frac{{\rm Vol}_{D-2}}{8 G}$. Unfortunately, it is not possible
to construct an algorithm that generates the sum over the triangulations of the
same volume. So the constant $\gamma$ is kept finite.
$\kappa_D(V,\kappa_{D-2})$ is chosen in such a way that
\begin{equation}
< N_D > = \sum_T \frac{1}{C_T}{\rm exp}(\kappa_{D - 2} N_{D - 2} - \kappa_{D}
N_D - \gamma (N_D - V)^2 ) N_D = V \label{kappa}
\end{equation}
This provides that the volume fluctuates around the required value $V$. The
fluctuations are of the order of $\delta V \sim \frac{1}{\sqrt{\gamma}}$. In
order to approach the model with constant $N_D$ we must keep $\frac{\delta
V}{V} << 1$. In practice we use $\gamma = 0.005$ for $V = 16000 , 32000$. So
$\frac{\delta V}{V} \sim 10^{-3}$.

For the numerical investigations we used Metropolis algorithm in its form
described in \cite{Catterall}. It is based on the following Markov chain. Each
step of the chain is the proposal of a deformation $T_i \rightarrow T_{f}$ of
the given triangulation $T_i$, which is accepted or rejected with the
probability ${\cal P}(T_i \rightarrow T_f)$ that satisfies the detailed balance
condition
\begin{equation}
{\rm exp}(-S(T_i)){\cal P}(T_i \rightarrow T_f) = {\rm exp}(-S(T_f)){\cal
P}(T_f \rightarrow T_i)
\end{equation}

The definition of the proposed deformations is based on the following idea. Let
us consider some closed $D$ - dimensional simplicial manyfold of the topology
of a $D$ - dimensional sphere. Then, if a connected piece of our original
triangulation is equal to a piece of this manyfold, we can replace it by the
remaining part of the given manyfold. Thus we obtain the deformed triangulation
with the same topology as the original one. Further, let us choose the boundary
$\partial s_{D+1}$ of a $D+1$ - dimensional simplex $s_{D+1}$ as the mentioned
above manyfold. There are $D+1$ opportunities to distinguish a piece of
$\partial s_{D+1}$ correspondent to $p$ - dimensional subsimplices of $s_{D+1}$
($ p = 0, ..., D$). The resulting deformation is called $(p, D - p)$ move
\cite{Catterall}.

It has been shown that via such moves it is possible starting from any
triangulation to reach a triangulation that is {\it combinatorially equivalent}
\footnote{Two triangulations are combinatorially equivalent if they have
subdivisions that are equivalent to each other up to the relabeling of
(sub)simplices. The subdivision $T_s$ of the given triangulation $T$ is another
triangulation such that the set of its vertices contains all vertices of $T$
and any simplex $s \in T_s$ either belongs to $T$ or belongs to some simplex of
$T$.} to an arbitrarily chosen other triangulation \cite{ergodicity}. This
property is called ergodicity. Due to this property starting from the
triangulation that approximates \cite{Geometry} the given Riemannian manyfold
it is possible to reach a triangulation that approximates almost any other
Riemannian manyfold. The exceptional cases, in which this is impossible, are
commonly believed not to affect physical results.

In practise we choose randomly the type of the move ($p \in \{ 0, ... , D \}$),
the simplex of a triangulation and its $p$ - dimensional subsimplex. After that
we check is there a vicinity of this $p$ - dimensional simplex that is
equivalent to the required piece of $\partial s_{D+1}$. If so, the suggested
move (and the correspondent subsimplex) is called legal and we proceed with
checking the possibility to perform the move \footnote{The move is not
geometrically allowed if the resulting  new simplex  already exists in the
given triangulation.}. If the move is allowed we accept or reject it with the
following probability:
\begin{equation}
p(T_i \rightarrow T_f) = \frac{1}{1+(1+\frac{N_D(T_f) -
N_D(T_i)}{N_D(T_i)}){\rm exp}(S(T_f) - S(T_i))} \label{P}
\end{equation}

In our calculations we start from the triangulation of minimal size that is
$\partial s_{D+1}$. Then we allow it to grow randomly until the volume reaches
the given value $V$. After that we proceed with the normal Metropolis process.
During this process the coupling $\kappa_D$ is self - tuned automatically in
order to get its required value satisfying (\ref{kappa}). This is done via
redefining it as $\kappa \rightarrow \kappa + 2 \gamma (<N_D>-V)$ after each
$10$ sweeps \footnote{One sweep is $V$ suggestions of  {\it legal} moves.}.

In order to make a check of our results, we have made two independent program
codes. The main program was written in $C^{++}$ using modern methods of object
- oriented programming.  The algorithm is partially based on the one described
in \cite{Catterall} and on the ideas suggested in \cite{deBakker}. The second
program was written in Fortran $77$ and was used for checking the main program
at small lattice sizes. The calculations were made within the parallel
programming environment using the computation facilities of Joint Supercomputer
Center (Moscow).

\section{Numerical results.}

We investigate the behavior of $10D$ model for $V = 8000$ and $V = 32000$. We
considered the values of $\kappa_D$ varying from $-0.1$ to $0.1$, where the
system is found to exist in two phases. The phase transition point is at
$\kappa_D \sim -0.03 $ at $V = 8000$ and at $\kappa_D \sim -0.01 $ at $V =
32000$.

The self - tuned value of $\kappa_{10}$ appears to be independent of $V$. In
accordance with large volume asymptotics obtained in \cite{Geometry} its
dependence upon $\kappa_8$ is linear with a good accuracy. The best fit is:
\begin{equation}
\kappa_{10} = 15.57(1) \kappa_{8} + 0.42(1)
\end{equation}

We investigate the following variables, which reflect the properties of the
triangulated manyfold.

1. The mean curvature carried by a simplex. The normalization is chosen in such
a way that it is defined as
\begin{equation}
R = \frac{ 4\pi N_{\rm bones}}{D(D+1) N_{\rm simplices} cos^{-1} (\frac{1}{D})}
- 1
\end{equation}

\begin{figure}
\begin{center}
\epsfig{figure=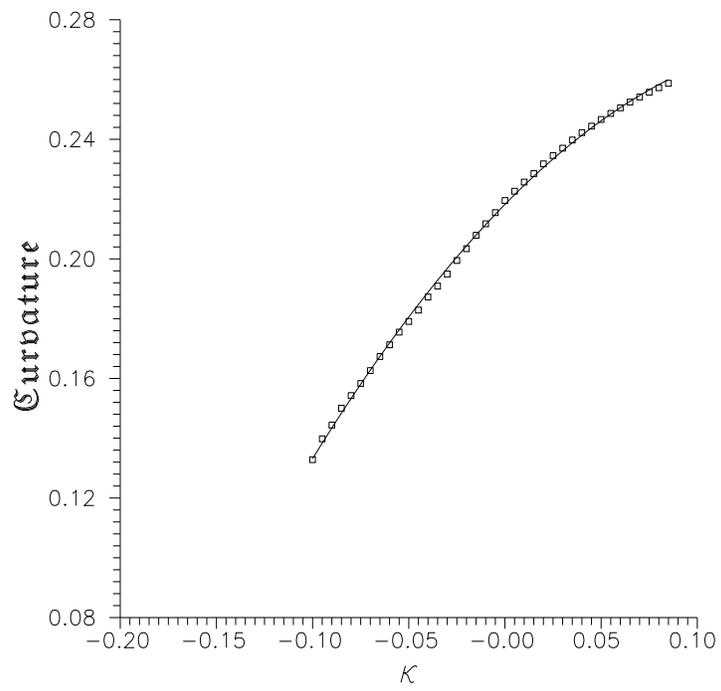,height=100mm,width=100mm,angle=0}
\caption{\label{fig.02} The curvature.
 \label{fig.2}}
\end{center}
\end{figure}

The results for $V = 8000$ and $V = 32000$ coincide with a good accuracy. They
are represented in the figure ~\ref{fig.02}. In contrast with the four
dimensional case (at the observed values of $\kappa_8$) $R > 0$ (in the four
dimensional case it becomes negative at the value of $\kappa_{D-2}$ close to
$0$). The best fit to the curve of the figure ~\ref{fig.02} is:
\begin{equation}
R = 0.218(1) + 0.654(2) \kappa_8 - 1.97(3) \kappa_8^2
\end{equation}

2. The geodesic distance between two simplices is the length of the shortest
path that connects them. The points of this path correspond to the simplices.
The links of the path correspond to the pairs of neighboring simplices. We
denote the geodesic distance between the simplices $u$ and $v$ by $\rho(u,v)$.
Let us fix a simplex $s$. Then the ball $B_R(s)$ of radius $R \in Z$ is
consisted of the simplices $u$ such that $\rho(u,s)$ is less than or equal to
$R$. The volume of the ball is defined as the number of simplices contained in
it.

One of the most informative characteristics of the triangulated manyfold is the
average volume ${\cal V}(R)$ of the balls (of the constant radius $R$) as a
function of this radius. In practise during each measurement we calculate
$V_R(s)$ for arbitrary chosen simplex $s$. Then we perform its averaging over
the measurements (separately for each $R$). ${\cal V}(R)$ becomes constant at
some value of $R$. This value is the averaged largest distance between two
simplices of the manyfold, which is also called the diameter $d$. We found that
for $V = 8000$ at $1 < R < \frac{d}{2}$ the dependence of ${\rm log} \,{\cal
V}$ on ${\rm log} \,R$ is linear. For $V = 32000$ the same takes place at $4 <
R < \frac{d}{2}$. The slopes give us the definition of the fractal (Hausdorff)
dimension of the manyfold: ${\rm log} \,{\cal V} = {\rm const} + {\cal D}{\rm
log} \,R$.

\begin{figure}
\begin{center}
 \epsfig{figure=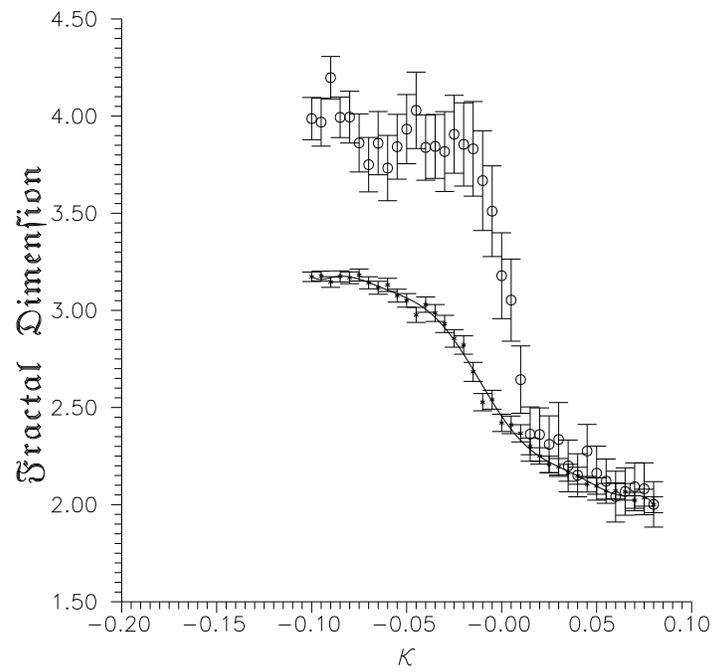,height=100mm,width=100mm,angle=0}
 \caption{\label{fig.04} The fractal dimension. Data for $V=8000$ is represented by
 the points and the solid
 line. Data for $V=32000$ is represented by the circles.
 \label{fig.4}}
\end{center}
\end{figure}

3. The mean fractal dimension ${\cal D}(V)$  of the manyfold for the observed
volumes $V$ as a function of $\kappa_{D-2}$ is represented on the
Fig.~\ref{fig.04}. The figure indicates that there is a phase transition at
critical $\kappa_{D-2} = \kappa_c(V)$. For the observed volumes $\kappa_c$
appears to be small and negative ($\kappa_c(8000) \sim -0.3$ and
$\kappa_c(32000) \sim -0.1$). For $\kappa > \kappa_c$ the fractal dimension is
close to two in accordance with the supposition that similar to the $4$ -
dimensional case the correspondent phase resembles branched polymers. For
$\kappa < \kappa_c$ ${\cal D}(8000) \sim 3.2$ and ${\cal D}(32000) \sim 4$.
This is in accordance with the expectations that this phase has a singular
nature and corresponds to ${\cal D}(\infty) = \infty$. Further we shall call
these two phases crumpled and elongated respectively. Our results give us
arguments in favor of view that the nature of these phases is similar to that
of the phases of the lower dimensional models. We must notice, however, that
the complete proof has not yet been obtained.

\begin{figure}
\begin{center}
 \epsfig{figure=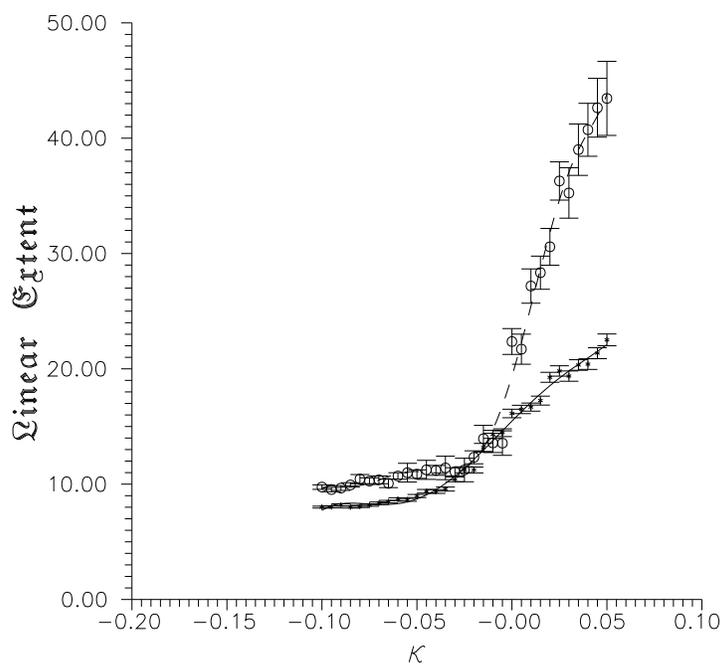,height=100mm,width=100mm,angle=0}
 \caption{\label{fig.01} The  linear extent.  Data for $V=8000$ is represented by
 the points and the solid
 line. Data for $V=32000$ is represented by the circles and the dashed line.
 \label{fig.1}}
\end{center}
\end{figure}

4. The linear size of the system can be evaluated as $V^{\frac{1}{\rm D}}$. So,
we expect that in the crumpled phase the diameter   $d \sim 10$ while in the
elongated phase   $d \sim 100$. These expectations are in accordance with the
direct measurements of the diameter as well as another parameter called linear
extent (see \cite{Catterall}). The linear extent ${\cal L}$ is defined as the
average distance between two simplices of the triangulation:
\begin{equation}
{\cal L} = \frac{1}{V^2}\sum_{u,v}<\rho(u,v)>
\end{equation}
Due to its construction $\cal L$ should be close to half a diameter. In our
measurements we calculate the linear extent using the slightly different
definition (that, anyway, should lead to the same result after averaging over
the measurements). Namely, we calculate
\begin{equation}
{\cal L}(s) = \frac{1}{V}\sum_{v}<\rho(s,v)> = \frac{1}{V}\sum_R R
(V_R(s)-V_{R-1}(s)).
\end{equation}
Performing the averaging over the measurements we obtain the required value of
the mean linear extent. Our results on $\cal L$ are represented in the figure
Fig.~\ref{fig.01}. One can see that the linear size of the manyfold is
increasing very fast in the elongated phase, while in the crumpled phase it
remains almost constant (and close to that of $3$, $4$ and $5$ dimensional
models \cite{Triangulation}). We found that the fluctuations of the linear
extent almost absent in the crumpled phase and are of the order of the mean
size of the manyfold in the elongated phase.

\section{Discussion.}

The force between two particles (of masses $m$) in the classical mechanics is
equal to
\begin{equation}
F = - \frac{G m^2}{R^2},
\end{equation}
where $R$ is the distance between them. Therefore negative gravitation coupling
$G$ and, correspondingly, negative $\kappa_{2}$ in the four - dimensional model
would correspond to repulsion instead of attraction. The same picture is, of
course, valid for the negative $\kappa_{D-2}$ in higher - dimensional models.
This means that the description of gravity in the dynamical triangulation
approach may appear only for the positive $\kappa_{D-2}$. The phase transition
in the pure Euclidean four - dimensional gravity is at $\kappa_2 \sim 1$ and
the model exists at physical couplings both in the crumpled and in the
elongated phases. In the ten dimensional model for the observed volumes the
phase transition point is $\sim - 0.01$. So at physical gravitational couplings
the system cannot exist
 in the crumpled phase.

The fractal dimension at $G > 0$ is found to be close to $2$.  This means that
in order to construct a dynamical triangulation model that corresponds to the
dynamics of Riemannian manyfold (of Euclidian metrics) of the dimension ${\cal
D} > 2$, it is necessary to add something to the pure Euclidean gravity.
However, the investigation of the latter seems sensible, because mechanisms
that appear in this model may play a role in more realistic models. From this
point of view the most interesting is the mechanism of compactification. We
started from the model, which has $\frac{D (D-1)}{2} = 45$ local degrees of
freedom correspondent to the independent components of the metric tensor $g$.
Finally we arrive (in the elongated phase) at the system, which has the fractal
dimension close to two. We expect that similar to the lower dimensional cases
this system can be effectively approximated by a tree graph. In the
correspondent branched polymer model \cite{Triangulation} the lengths of the
linear pieces of the graph and the way of their gluing together are the
dynamical variables. What does then happen with the remaining degrees of
freedom? They correspond to the small fluctuations of the pieces of the
original manyfold represented by the links of the graph. Those fluctuations,
obviously, become the internal degrees of freedom living on the polymer. This
dimensional reduction should, in principal, resemble the dimensional reduction
of the Kaluza - Klein models \cite{Kaluza}. However, the concrete realization
of the mechanism has to be investigated. We expect that more realistic
realizations of the dynamical triangulation approach should possess the
dimensional reduction, which may be similar to that of the pure Euclidean
gravity.

The investigation of the $5$ - dimensional model (see \cite{5D}) indicates that
it has a complicated phase structure. Namely, at $\kappa_{D-2} < -5$ several
different vacua appear. And the system possesses tunnelling between them. In
our research of the $10$ - dimensional gravity we restricted ourselves by the
ranges of $\kappa_8 \in (-0.1, 0.1)$, where the phase transition between
crumpled and elongated phases is observed. However, taking into account the
above mentioned property of the $5D$ model we expect that  the phase structure
of the considered model is, probably, not limited by the observed two phases.
Another open question is the order of the observed phase transition. Although
we have some indications that it is of the first order, the complete
investigation of the subject has not been performed.

\section{Conclusions.}

In this paper we report our results on the numerical investigation of the $10$
- dimensional Euclidean quantum gravity in the framework of the dynamical
triangulation approach. Our summary is as follows.

1. The considered model contains  the phases that we call crumpled and
elongated. The arguments in favor of view that those phases resemble the
correspondent phases of lower dimensional models were obtained.

2. For the observed volumes the phase transition between the mentioned phases
corresponds to small negative values of $\kappa_{8}$. Therefore at the
"physical" positive gravitational coupling $G \sim \frac{1}{\kappa_8}$ the
model does not contain the crumpled phase contrary to the lower dimensional
cases, where critical value of $\kappa_{D-2}$ is positive. This also means that
the natural state of the collection of $10$ D simplices (that appears when the
Einstein - Hilbert action is turned off and $\kappa_8 = 0$) belongs to the
elongated phase (at least for the observed volumes). This is in the contrast
with the behavior of the low dimensional models: at $D = 3,4$ and $\kappa_{D-2}
= 0$ the systems exist in the crumpled phase.

3. The average scalar curvature appears to be positive at the observed
gravitational couplings ($\kappa_8 \in (-0.1, 0.1)$) in contrast with, say, the
$4$ - dimensional case, where it becomes negative at $\kappa_2$ close to $0$.

4. The fractal dimension of the model is close to $2$ in the elongated phase
for both considered volumes ($V = 8000$ and $V=32000$). In the crumpled phase
the fractal dimension appears to be $\sim 3.2$ for $V = 8000$ and ${\cal D}
\sim 4$ for $V = 32000$.

5. The linear size of the system in the crumpled phase appears to be close to
that of the lower dimensional models (the linear extent is $\sim 10$). In the
elongated phase the mean linear extent and its fluctuations appear to be
sufficiently larger than in the crumpled phase.

6. It is expected that the considered phase transition is of the first order.
However, the investigation of this subject has not been performed. So the
question about the order of the phase transition remains open.

\section{Acknowledgments.}

We are grateful to E.T Akhmedov, B.L.G. Bakker, M.I.Polikarpov and J.Smit for
useful discussions. We are very obliged to J.Ambjorn and S.Catterall who have
sent us the dynamical triangulation (Fortran and C) program codes. We consulted
these codes several times during the elaboration of our own program codes. We
are very obliged to the staff of the Joint Supercomputer Center at Moscow and
especially to A.V. Zabrodin for the help in computations on supercomputer MVS
1000M. This work was partly supported by Federal Program of Russian Ministry of
Industry, Science and Technology N 40.052.1.1.1112 and by the grants RFBR
03-02-16941, RFBR 04-02-16079, RFBR 02-02-17308, RFBR 01-02-117456, RFBR-DFG
03-02-04016, INTAS 00-00111 and CRDF award RP1-2364-MO-02.

\end{document}